\documentstyle[aps,prbbib]{revtex}  
\begin{document}
\draft
\title{Spin-polarized transport and Andreev reflection in 
semiconductor/superconductor hybrid structures}
\author{Igor \v{Z}uti\'c \cite{igor} and S. Das Sarma}
\address{Department of Physics, 
University of Maryland at College Park, College Park, Maryland 20742-4111} 
\maketitle
\begin{abstract}
We show that spin-polarized electron transmission across  
semiconductor/superconductor (Sm/S) hybrid structures 
depends sensitively on the degree of spin polarization 
as well as  the strengths
of potential and spin-flip scattering at the interface. 
We demonstrate that increasing the 
Fermi velocity mismatch in the Sm and S regions can
lead to enhanced junction transparency in the presence of 
spin polarization. We find that the Andreev reflection amplitude
at the superconducting gap energy is a robust 
measure of the  spin polarization magnitude, being independent of
the strengths of potential and spin-flip scattering and the Fermi
velocity of the superconductor.
\end{abstract}

We study theoretically spin-polarized carrier transport across 
semiconductor/ superconductor (Sm/S) hybrid structures. 
Our motivation arises from the increasing current interest in spin transport
in electronic materials\cite{prinz} and the possibility of using spin degrees 
of freedom in nanoelectronic devices.
The new terminology ``spintronics,'' coined as an
alternative to the charge based conventional electronic technology, has
emerged encompassing 
diversified efforts to understand and exploit various aspects
of spin dynamics,\cite{jaro} which could lead to the design of 
potentially novel devices.
Many of the recent advances have focused on studying Sm and
hybrid Sm structures as the likely candidates for spintronic
devices.\cite{das,aws,john}
Although the well established growth and materials fabrication
techniques for semiconductors coupled with their
tunable electronic properties, such as carrier
density and Fermi velocity,  provide important advantages over other
materials,  it is nevertheless essential to  demonstrate the feasibility of 
creating and controlling spin-polarized carriers\cite{his} in semiconductors,
if they are going to be utilized in proposed spintronic devices. 
In addition to the optically
created spin-polarized electrons,\cite{aws,hag} 
it is now possible to employ several other
methods to produce spin-polarized carriers in semiconductors, such as, 
the direct injection from a ferromagnet,\cite{ham}
using a novel class of ferromagnetic semiconductors\cite{ohno} 
based on Mn-doped GaAs, and 
utilizing  the Bychkov-Rashba effect.\cite{silva} 

For the  operation of hybrid Sm structures it is
often desirable to achieve high junction transparency
(i.e. a large  transmission coefficient) at the interface
of a semiconductor with other materials. 
The junction  transparency, at the semiconductor 
interface with a metal, is typically 
limited by a native Schottky\cite{wolf}  barrier.
Recently, in the context of unpolarized spin transport, novel fabrication
techniques were developed\cite{belt} to suppress the Schottky barrier, 
providing, as a consequence, ohmic contacts and ballistic transport,
across Sm/metal interfaces.

Our aim in this paper, motivated by the recent advances in hybrid 
Sm  structures, is  to investigate and quantify
their  degree of carrier spin polarization and the junction transparency.
For this purpose we study 
Sm/S structures\cite{lamb,rand} 
and the process of Andreev reflection\cite{and} (governing the low 
bias transport) which is very sensitive to both the 
degree of the spin polarization and the junction transparency.
Such sensitivity can be explained by noticing that the Andreev reflection is
a two-particle process. An incident electron of spin 
$\sigma=\uparrow,\downarrow$ on a Sm/S interface is reflected as a 
hole belonging to the opposite spin subband $\overline{\sigma}$, back
to the Sm region while a Cooper pair is transfered to the superconductor.
The probability for Andreev reflection at low bias voltage is thus
related to the square of the normal state transmission coefficient and 
has a stronger dependence on the junction transparency than the ordinary
single particle tunneling.
For  spin-polarized carriers, with  different populations in two spin
subbands, only a fraction of the incident electrons from a majority subband 
will have a  minority subband partner in order to be  Andreev
reflected. Specifically, in the limit of full spin polarization
Andreev reflection is completely suppressed. In the superconducting
state, for an applied voltage smaller than the superconducting gap,
single particle tunneling is not allowed in the S region and the 
modification of the
Andreev reflection amplitude by spin polarization or junction transparency
will be  manifested  in  transport quantities such as  the differential 
conductance or current-voltage characteristics of  Sm/S junction.

In  previous work on Sm/S structures, only  spin-unpolarized 
transport\cite{lamb,rand}
was considered, and the spin-dependent Andreev reflection has only been 
addressed in the context of ferromagnet/S junctions.\cite{vas2}
However, in the latter case, calculations were performed  assuming 
the equality of the effective masses\cite{been,zv} in the two regions
across the interface.
For Sm/S 
structures, which have a typical ratio of the effective masses $\sim 10$, such 
an assumption is clearly incorrect, and 
therefore we generalize to a situation with both unequal effective masses and
Fermi velocities\cite{dif}. 

We consider transport of injected spin-polarized carriers by choosing
a geometry where the Sm occupies $x<0$ region and is separated by the
flat interface, at $x=0,$ from the S region, at $x>0$. 
We solve the Bogoliubov-de Gennes (BdG) equations\cite{zv,deg} for
a  Sm/S junction assuming specular reflection and 
translational invariance 
parallel to the interface ($y-z$ plane),
implying the conservation of the parallel component of the wavevector 
${\bf k}_\|$ 
in the scattering processes.
At the interface we consider both potential (non-spin-flip)
 and  spin-flip scattering. For an electron with spin $\sigma$, 
incident from the Sm region, the allowed scattering
processes and the corresponding amplitudes include:\cite{btk}  
Andreev reflection, $a_\sigma$, ordinary (normal) reflection, $b_\sigma$, 
transmission as an electronlike quasiparticle, $c_\sigma$, the transmission 
with the branch crossing as a holelike quasiparticle, $d_\sigma$, together with
their spin-flip counterparts, denoted by 
$a_{f\sigma}$, $b_{f\sigma}$, $c_{f\sigma}$,
$d_{f\sigma}$. For example, the amplitude $b_{f\sigma}$ will correspond to the
reflection of the incident electron accompanied by a spin-flip.
The appropriate angle for each scattering trajectory is determined by an
analogue of Snell's law.\cite{zv}
In the presence of spin-flip scattering BdG equations can no longer be 
decoupled\cite{been,zv} into spin up/down sectors. Instead, we look for 
four dimensional solutions of the form 
$\Psi_\sigma=(\phi_\sigma, \chi_{\overline{\sigma}},\phi_{\overline{\sigma}},
\chi_\sigma)^T$, where 
$\phi_\sigma$, $\chi_{\overline{\sigma}}$, are the electronlike quasiparticle
(ELQ) and the holelike quasiparticle (HLQ) amplitudes, respectively. 
We consider  ballistic transport in Sm and S regions and model the interfacial
scattering by $U({\bf r})=\tilde{U} \delta(x)$, where $\delta(x)$ is
the Dirac delta function and $\tilde{U}$ is a 
4 $\times$ 4 matrix with the 
$diag(H_\uparrow,H_\uparrow,H_\downarrow,H_\downarrow)$
representing the (possibly) spin-dependent strength\cite{fil} 
of potential scattering, while
the off-block-diagonal elements govern the spin-flip scattering.
From the
 conservation of ${\bf k}_\|$, and by writing 
$\Psi_\sigma({\bf r})\equiv e^{i {\bf k}_{\|\sigma}\cdot
 {\bf r}}\psi_\sigma(x)$,  the problem becomes an effectively one-dimensional 
four-component interface scattering problem.
Within the effective mass approximation, the
boundary conditions ensuring probability conservation  are
\begin{eqnarray}
\psi_{1\sigma}(0)=\psi_{2\sigma}(0), \quad
\frac{\hbar}{m_1}\partial_x \psi_{1\sigma}(0)
=\frac{\hbar}{m_2}\partial_x \psi_{2\sigma}(0)
-\frac{1}{\hbar}\tilde{U}\psi_{1\sigma}(0),
\label{bc}
\end{eqnarray}
where subscripts $1$, $2$ corresponds to the quantities in the Sm and S
regions, respectively.
We use  a step function form of the pair potential 
$\Delta(x)=\Delta e^{i\phi} \Theta(x)$,\cite{rand} where $\Delta$ is the
superconducting gap.
We  employ the  Andreev approximation,\cite{rand} 
valid for $\epsilon,\Delta \ll E_F$, where $\epsilon$  
and $E_F$ are  the excitation and the Fermi energy, respectively. The  
solution in the Sm region, for spin $\sigma$ 
electrons incident at an angle $\theta$ with the normal to the interface is
\begin{eqnarray}
\psi_{1\sigma}(x)=
e^{i  k_{1\sigma}  x} 
\left[\begin{array}{c} \delta_{\uparrow \sigma}
 \\ 0 \\ \delta_{\downarrow \sigma} \\ 0 \end{array} \right]
+a_\sigma e^{i k_{1\overline{\sigma}}x} 
\left[\begin{array}{c} 0 \\ \delta_{\uparrow \sigma}
 \\ 0 \\ \delta_{\downarrow \sigma} \end{array} \right]
+a_{f\sigma} e^{i k_{1\sigma} x} 
\left[\begin{array}{c} 0 \\ \delta_{\downarrow \sigma}
 \\ 0 \\ \delta_{\uparrow \sigma} \end{array} \right]
+b_\sigma e^{-i k_{1\sigma} x} 
\left[\begin{array}{c} \delta_{\uparrow \sigma}
 \\ 0 \\ \delta_{\downarrow \sigma} \\ 0 \end{array} \right]
+b_{f\sigma} e^{-i k_{1\overline{\sigma}}x} 
\left[\begin{array}{c} \delta_{\downarrow \sigma}
 \\ 0 \\ \delta_{\uparrow \sigma} \\ 0 \end{array} \right],
\label{smup}
\end{eqnarray}
where by $k_{1\sigma}$, $k_{1\overline{\sigma}}$, 
we denote, generally different $x$-components of ${\bf k}_1$,
and $\delta_{\uparrow,\downarrow \sigma}$ is the Kronecker symbol.
Analogously, solutions in the S region can be explicitly written, and 
consequently all the 
scattering amplitudes analytically determined from  Eq. (\ref{bc})
using the explicit forms of the wavefunctions.
The conservation of the parallel component implies
 ${\bf k}_{\|\sigma} \equiv {\bf k}_{i \| \sigma}
= {\bf k}_{1 \| \overline{\sigma}}$, $\sigma=\uparrow,\downarrow$, $i=1,2$,
and within the Andreev approximation, conservation of energy requires
${\bf k}^2_{1F\sigma}=k^2_{1\sigma}+k^2_{\| \sigma}=k^2_{1\overline{\lambda}}+
k^2_{\| \lambda}$, ${\bf k}^2_{2F}=k^2_{2\sigma}+k^2_{\| \sigma}$,
$\lambda=\downarrow,\uparrow$, where 
${\bf k}_{1F\sigma}$,
${\bf k}_{2F}$ are the Fermi wavevectors in the Sm and S regions, 
respectively.
We assume that in the presence of spin polarization, $k_{1F\uparrow} \ge
k_{1F \downarrow}$ holds, as a consequence of the splitting, $h_0$,
of the spin subbands
with corresponding bandwidths $E_{F\sigma}=E_F+\rho_\sigma h$,
where $\rho_\sigma=\pm1$.  
 Consequently, there can exist a range for $\theta$
(less than the angle of total reflection) where $k_{1  \overline{\uparrow}}$
is purely imaginary while there is still nonvanishing transmission across the
junction.\cite{zv} For this purpose, we extend the above notation to the
$x$-components of velocity and define
 $v'_{1 \overline{\uparrow}}\equiv Re(v_{1 \overline{\uparrow}})$. 
In contrast, from $k_{1F\uparrow} \ge
k_{1F \downarrow}$, it follows that  
 $v'_{1 \overline{\downarrow}}$ and $v_{1 \overline{\downarrow}}$
can be interchanged since the imaginary part would only arise
at $\theta$ greater than the critical angle for total reflection.

To obtain the differential conductance and related quantities of interest
in spin-polarized transport studies,
we modify the BTK method,\cite{btk} along the line of the 
Landauer-B\"{u}ttiker\cite{land}  scattering
formalism in the normal state transport.
Employing the conservation of probability current\cite{noz} it is necessary to
calculate only the amplitudes from Sm region, giving the differential
conductance 
\begin{eqnarray}
G^{{\rm Sm/S}}=\frac{e^2}{h} \sum_{{\bf k}_{\| \sigma},\; \sigma}
\int^\infty_{-\infty}
[1+\frac{v'_{1\overline\sigma}}{v_{1\sigma}}(|a_\sigma|^2-|b_{f\sigma}|^2)
+|a_{f\sigma}|^2-|b_\sigma|^2] [f(\epsilon-e V)-f(\epsilon)] d \epsilon,
\label{gt}
\end{eqnarray}
where $V$ represents the bias voltage, and $f(\epsilon)$
is the Fermi function,
which introduces temperature dependence in the problem.
For our purposes, 
it is sufficient to focus on the low temperature regime where
the Fermi function difference in Eq. (\ref{gt})
can be replaced by the Dirac delta function 
implying $\epsilon=eV$. The substitution
$[1+\frac{v'_{1\overline\sigma}}{v_{1\sigma}}(|a_\sigma|^2-|b_{f\sigma}|^2)
+|a_{f\sigma}|^2-|b_\sigma|^2]$ $\rightarrow$
$[1-\frac{v'_{1\overline\sigma}}{v_{1\sigma}}(|a_\sigma|^2+|b_{f\sigma}|^2)
-|a_{f\sigma}|^2-|b_\sigma|^2] \rho_\sigma$ in Eq. (\ref{gt}), 
gives an explicit expression for the spin conductance. Noticing that the
latter term is proportional to the quasiparticle current, 
we can immediately infer that the subgap spin conductance will
vanish as there is no quasiparticle tunneling below the 
superconducting gap.

In the absence of spin-flip scattering $a_{f\sigma}$  and $b_{f\sigma}$
vanish, and for the remaining amplitudes we obtain
\begin{eqnarray}
a_\sigma=\frac{4 v_{1\sigma} v_{2\sigma} \Gamma}
{( v_{1\sigma} v_{1\overline\sigma}+v^2_{2\sigma}+4H_\sigma^2/\hbar^2
+2i( v_{1\sigma}-v_{1\overline\sigma}) H_\sigma/\hbar)(1-\Gamma^2)
+v_{2\sigma}( v_{1\sigma}+ v_{1\overline\sigma})(1+\Gamma^2)},
\label{as}
\end{eqnarray}
\begin{eqnarray}
b_\sigma=\frac
{(v_{1\sigma} v_{1\overline\sigma}-v^2_{2\sigma}-4H_\sigma^2/\hbar^2
-2i( v_{1\sigma}+v_{1\overline\sigma}) H_\sigma/\hbar)(1-\Gamma^2)
+v_{2\sigma}( v_{1\sigma}- v_{1\overline\sigma})(1+\Gamma^2)}
{( v_{1\sigma} v_{1\overline\sigma}+v^2_{2\sigma}+4H_\sigma^2/\hbar^2
+2i( v_{1\sigma}-v_{1\overline\sigma}) H_\sigma/\hbar)(1-\Gamma^2)
+v_{2\sigma}( v_{1\sigma}+ v_{1\overline\sigma})(1+\Gamma^2)},
\label{bs}
\end{eqnarray}
where $\Gamma$
is related to the BCS coherence factors and is given by
$\Gamma
=[\epsilon-i\sqrt{\Delta^2-\epsilon^2}]/\Delta$
for $|\epsilon|\leq \Delta$, and
 $\Gamma=[\epsilon-sign(\epsilon)\sqrt{\epsilon^2-\Delta^2}]/\Delta$
for $|\epsilon|> \Delta$.
Using the  Eq. (\ref{as}) and (\ref{bs}) 
it is interesting to compare the effective transmission coefficients
for Sm/normal metal (Sm/N) and Sm/S junctions at zero bias 
\begin{eqnarray}
T^{\rm {Sm/N}}_\sigma(\theta)=\frac{4 v_{1\sigma} v_{2\sigma}}
{(v_{1\sigma}+v_{2\sigma})^2+4H^2_\sigma/\hbar^2}, \quad
T^{{\rm Sm/S}}_\sigma(\theta)=\frac{4 v_{1\sigma} v'_{1\overline{\sigma}}
 v^2_{2\sigma}}
{(v_{1\sigma} v'_{1\overline{\sigma}}+v^2_{2\sigma}+4H^2_\sigma/\hbar^2)^2
+4 (v_{1\sigma}-v'_{1\overline{\sigma}}) H^2_\sigma/\hbar^2}.
\label{smn}
\end{eqnarray}
For normal
incidence, $\theta=0$ 
($v_{1\sigma} v'_{1\overline{\sigma}}= v_{1F\uparrow} v_{1F\downarrow}$,
 $v_{2\sigma}=v_{2F}$), and
in the absence of potential scattering $H_\sigma=0$: therefore
in order to attain  perfect transparency ($T=1$), it is necessary in 
both Sm/N
and unpolarized Sm/S 
junctions for the Fermi velocities
$v_{1F\uparrow,\downarrow}$, $v_{2F}$ to be equal.
However, in the presence of finite  spin polarization in Sm/S junctions it
is possible to have perfect transparency even when all the 
Fermi velocities differ, satisfying  
$v_{2F}=\sqrt{v_{1F\uparrow} v_{1F\downarrow}}$.\cite{why}
There is also a broader
regime, $H_\sigma$, $\theta\neq0$, which can be determined using Eq. (\ref{as}), (\ref{bs}), 
where finite  spin polarization ($v_{1F\uparrow} \neq v_{2F\downarrow}$)
enhances  junction transparency as compared to the unpolarized case. 

To investigate 
 the effect of the variable degree of spin polarization we  first consider
a  geometry which models the in-plane transport of a two-dimensional 
electron gas (2DEG) which  has a quasi one-dimensional interface, of length $L$
 with a superconductor.  For illustration, we choose parameters  $m_2/m_1=15$, 
and $k_{2F}/k_{1F}=20$,  corresponding approximately to 
Al/GaAs interface with a Sm 2DEG  carrier density 
$n_s\sim 10^{13} {\rm cm}^{-2}$, and thickness
$\lesssim 70$ nm with occupation only of the first quantized level.\cite{sim}
We focus  on the  normalized conductance $G_2=G^{{\rm Sm/S}}/G_{2N}$,
where $G_{2N}=e^2 L(k_{1F\uparrow}+k_{1F\downarrow})/h\pi$ is the normal
state conductance, and 
$G^{{\rm Sm/S}}$ is given in Eq. (\ref{gt}). The summation
in Eq. (\ref{gt}), can be replaced by $\propto \int d \theta \cos \theta$,
favoring the forward scattering channel.
We parameterize potential scattering\cite{btk} by 
$Z_\sigma=H_\sigma/(v_{1F}v_{2F})^{1/2}$
and spin-flip 
scattering by $F=H_F/(v_{1F}v_{2F})^{1/2}$, where  $H_F$ is
the phenomenological parameter in matrix $\tilde{U}$, 
which determines the strength of 
$\phi_\sigma$ $\rightarrow$ $\phi_{\overline{\sigma}}$ and  
$\chi_\sigma$ $\rightarrow$ $\chi_{\overline{\sigma}}$
scattering. In this approach, our intention is to  investigate some of the 
qualitative features  of spin-flip scattering, without 
specifying its precise origin.
To model the subband spin-splitting and the
related degree of spin polarization, we define
the parameter $X\equiv h_0/E_f$. In Fig. \ref{f1}, we give
the calculated $G_2(e V/\Delta)$ for various degrees 
of spin polarization (including strong polarization, currently not 
attainable in the hybrid semiconductor structures), in the absence
of interfacial scattering, $Z_\sigma=F=0$. 
In the inset we show the effect 
of interfacial scattering at a fixed spin polarization.
While individually  potential and  spin-flip 
scattering each tends to suppress  the conductance below and above the gap, 
their combined effect may surprisingly 
lead to the enhancement of subgap conductance.
Conductance amplitude for  fixed $X$, at the bias voltage
of the superconducting gap, depends (using the conservation of
probability current) only on amplitudes $a_\sigma$ and $a_{f\sigma}$.
From Eq. (\ref{as}) and by noticing that $\Gamma=1$, at $eV=\Delta$, 
it follows that $a_\sigma$ depends on the ratio of $v_{1\sigma}$ and
$v_{1\overline{\sigma}}$, i.e., on the subband spin-splitting only,  while it
can be shown that $a_{f\sigma}$ vanishes identically in this situation.

We next consider  the geometry for a bulk Sm/S  or a 2DEG/S junction with a 
two-dimensional interface, parallel to the 2DEG. 
Calculating the conductance from Eq. (\ref{gt}) entails summation over
the two ${\bf k}_\|$ components which could be replaced
 by $\propto \int d \theta \cos \theta \sin \theta$.
We define $G_3= G^{{\rm Sm/S}}/G_{3N}$ where
$G_{3N}=e^2 A (k^2_{1F\uparrow}+k^2_{1F\downarrow})/h \pi=  
k^2_F/h \pi$ is the Sharvin conductance in the normal state and  $A$ is
the contact area of the Sm/S interface. 
In Fig. \ref{f2}, we use the same parameters as in the previous
figure, omitting  results for $X>0.6$. 
For a bulk Sm region, the chosen ratio of $k_{2F}/k_{1F}$ can be achieved
with a large surface doping.\cite{belt}
The angular dependence,
of the combination of scattering amplitudes in $G_2$ and $G_3$ is identical.
There are, however, differences  between the figures, 
including the smaller subgap conductance 
in the latter case,  which can be explained by the angular
integration, weighting more contributions near $\theta=\pi/4$,
rather than $\theta=0$ region (as in the former case). 
The inset in Fig. \ref{f2} shows  that the conductance peak is no longer
exactly at the superconducting gap, as compared to the one in the
previous figure. The maximum value depends only weakly 
on the presence of potential and spin-flip scattering and together with
other features in $G_3$ can be used
to estimate the degree of spin polarization from the conductance data. 
The effect of combined spin-flip and potential
scattering is more pronounced in Fig. \ref{f2} and
is qualitatively different from
Fig. \ref{f1}, in particular the double conductance peak 
in Fig. \ref{f2} has no analog in Fig. \ref{f1}.

In summary, we have shown that the low temperature transport
properties in  Sm/S structures may serve as a  sensitive and 
quantitative probe 
for determining the degree of spin polarization and 
the strength of interfacial scattering. 
The high Sm/S junction transparency achieved in recent fabrication techniques, 
can be further enhanced with spin-polarized carriers. Experimental studies 
of Sm/S junctions should provide an important testing ground 
for studying the  feasibility of spintronic devices
based on  hybrid semiconductor structures. Because of
the various simplifying approximations (e.g. specular scattering
with conserved parallel component of wavevectors,
delta function model for interface 
scattering, step function approximation for the superconducting gap function,
etc.) our results should be taken as of qualitative rather
than of quantitative validity.
However,  all of these approximations are nonessential,
made for the sake of analytical simplicity and may be improved
upon in future numerical treatments of the problem  when experimental
data on spin-polarized transport in Sm/S hybrid structures becomes available.

We thank J. Fabian and I.I. Mazin for useful discussion. 
This work was supported by the U.S. ONR and DARPA. 
%

%
\begin{figure}
\caption{Normalized conductance, $G_2(eV/\Delta)$,
curves from top to bottom represent  
X=0,0.1,0.2,0.3,0.4,0.5,0.6,0.7,0.8,0.9,0.95 at $Z_\sigma=F=0$. 
The inset shows $X=0.4$ results, the upper
four curves (from top to bottom at zero bias) have $Z_\sigma$, $F$ values
given by (0,0), (0.5,0.5), (0.5,0.25), (0.5,0.25), and (0,0)
for the bottom curve, corresponding to $k_{2F}/k_{1F}=70$
and  $n_s\sim 10^{12}{\rm cm}^{-2}$.} 
\label{f1}
\end{figure}
\begin{figure}
\caption{Normalized conductance, $G_3(eV/\Delta)$.
Curves from top to bottom correspond to 
$X=0,0.1,0.2,0.3,0.4,0.5,0.6$, all the other
parameters and ordering  are the same as in 
Fig. \protect{\ref{f1}} and its inset.}
\label{f2}
\end{figure}

\begin{references}
\bibitem{dif} They are inequivalent
since carrier densities across the interface are in general different.
\bibitem{fil} $H_\uparrow \neq H_\downarrow$ can describe magnetically
active interface and spin-filtering.
\bibitem{why} This form is a consequence of different flux factors,
the ratios of Fermi velocities, which appear in the transmission coefficient.
\bibitem{sim} The model is simplistic and is mainly being 
used for the purpose
of illustration, it can be easily improved upon to correspond to
experimental systems.
\end{references}
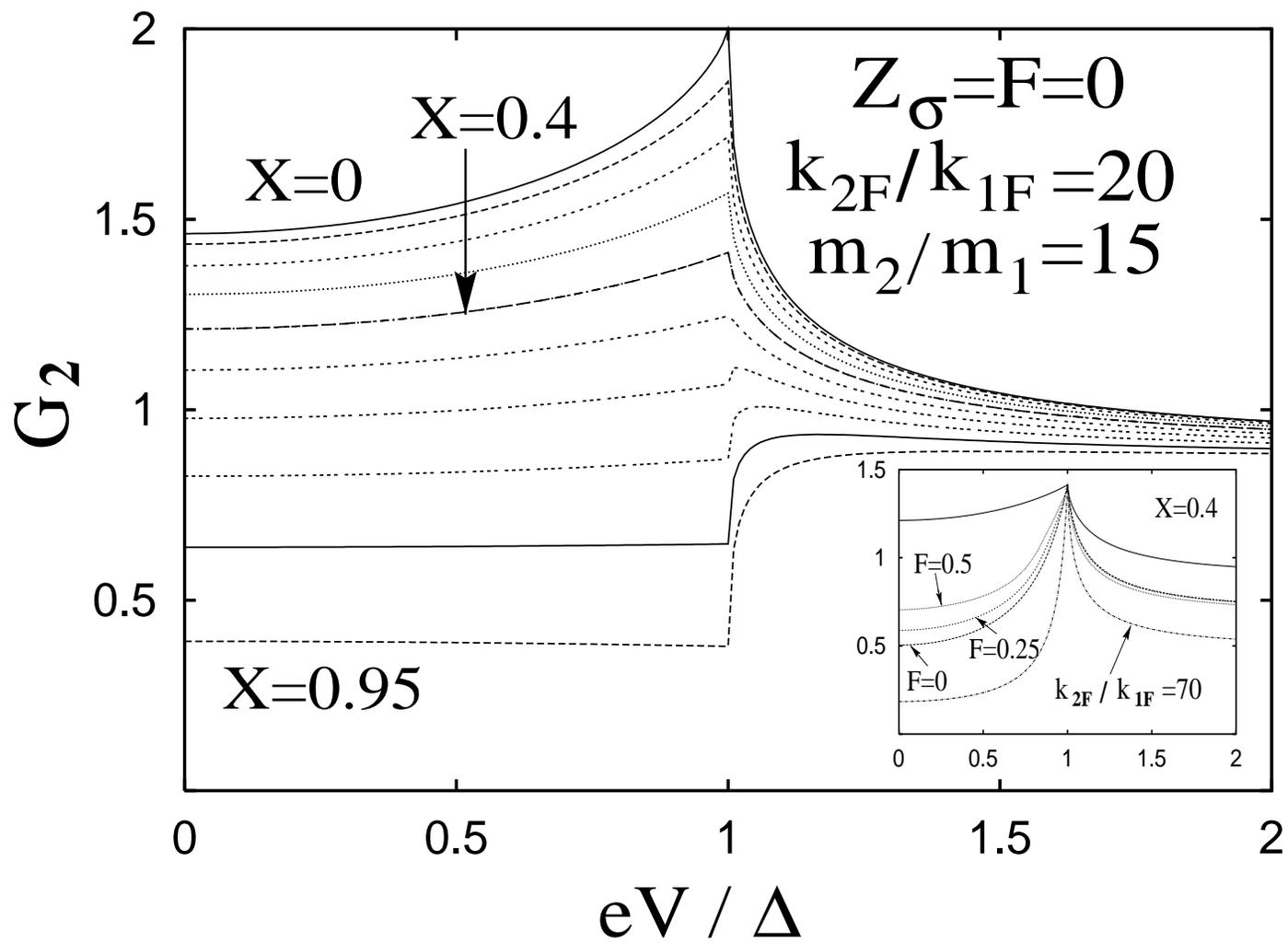
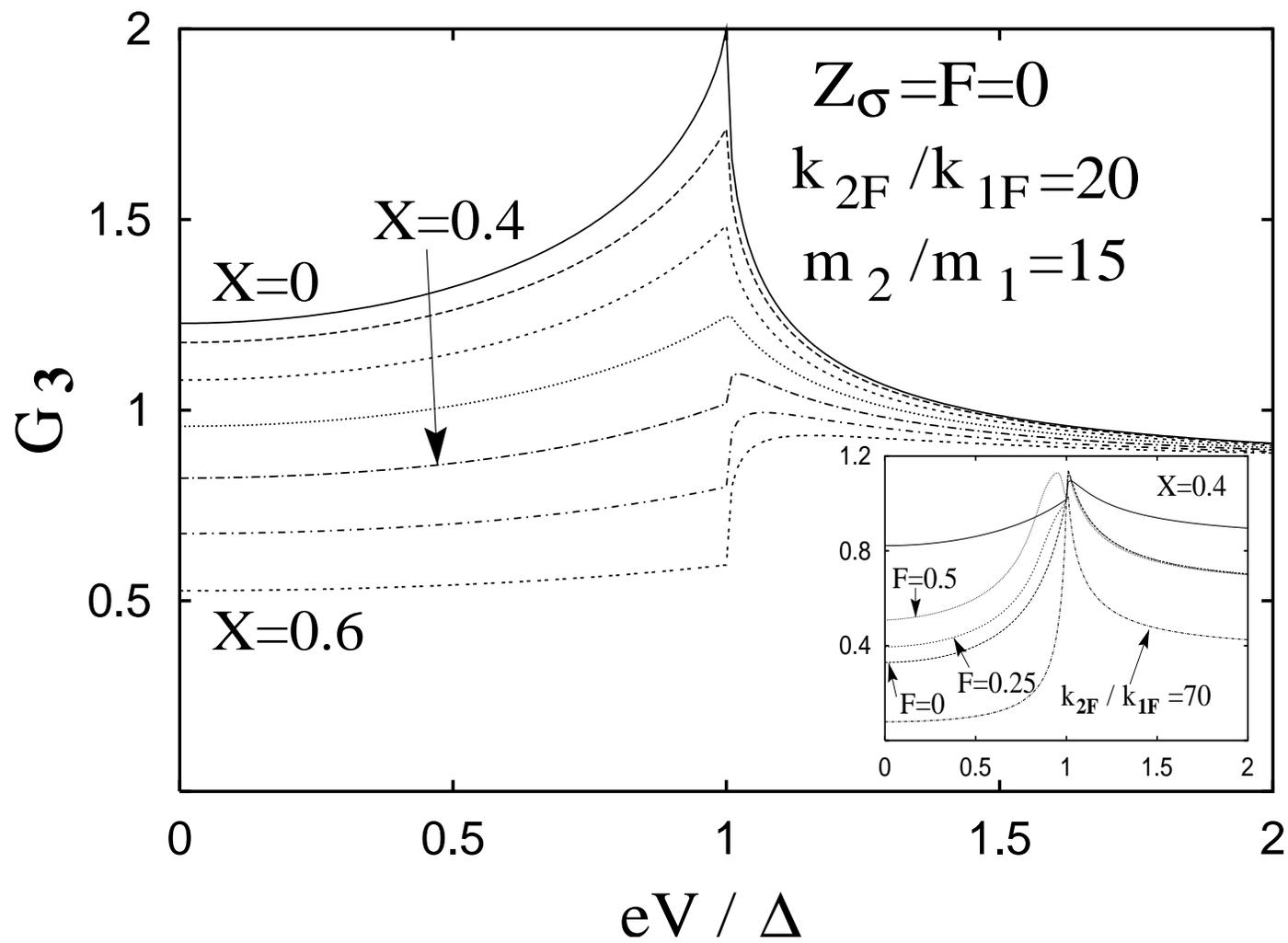
\end{document}